\documentclass[graybox]{svmult}

\usepackage{helvet}         
\usepackage{courier}        
\usepackage{type1cm}        
\usepackage{makeidx}         
\usepackage{graphicx}        
\usepackage{multicol}        
\usepackage[bottom]{footmisc}

\usepackage{xcolor}
\usepackage{graphicx} 
\usepackage{booktabs} 
\usepackage{subfig}
\usepackage[english]{babel}
\usepackage{amsfonts}
\usepackage{amssymb}
\usepackage{amstext}
\usepackage{amsmath}
\usepackage{amsmath,bm}

\usepackage{multicol}
\usepackage{mwe,tikz}\usepackage[percent]{overpic}

\newcommand\mba{\mathbf{a}}
\newcommand\mbb{\mathbf{b}}
\newcommand\bee{\begin{equation*}}
\newcommand\eee{\end{equation*}}
\def\be{\begin{equation}}
\def\ee{\end{equation}}
\def\ba{\begin{eqnarray}}
\def\ea{\end{eqnarray}}

\newcommand\dphi{\dot{\phi}}
\newcommand\eb{b_0 e^{-\alpha\phi}}

\begin{document}

\title*{The role of the slope in the the multi-measure cosmological model}
\author{Denitsa Staicova}
\titlerunning{The slope of the multi-measure model}
\institute{Denitsa Staicova \at Institute for Nuclear Research and Nuclear Energy \\Bulgarian Academy of Sciences, Sofia 1784, Tsarigradsko shosse 72, Bulgaria, \email{dstaicova@inrne.bas.bg}}
%
%
\maketitle

\abstract{In this work, we report some results on the numerical exploration of the model of Guendelman-Nissimov-Pacheva. This model has been previously applied to cosmology, but there were open questions regarding its parameters. Here we demonstrate the existence of families of solutions on the slope of the effective potential which preserve the duration of the inflation and its power. For this solutions, one can see the previously reported phenomenon of the inflaton scalar field climbing up the slope, with the effect more pronounced when starting lower on the potential slope. Finally we compare the dynamical and the potential slow-roll parameters for the model and we find that the latter describe the numerically observed inflationary period better. 
}

\section{Cosmology today}
Some of the most defining features of the Universe we live in are that it is isotropic, homogeneous and flat. They have been confirmed to great precision by cosmological probes (WMAP, Planck). Another important observation is that the universe is currently expanding in an accelerated way (confirmed by the data from SNIa and the Cepheids) which requires the introduction of dark energy. A model which describe all of those fundamental properties is the $\Lambda-CDM$ model, in which different components of the energy density contribute to the evolution of the universe as different powers of the scale factor. 

Explicitly, in the Friedman-Lemaitre-Robertson-Walker (FLRW) metric $\tilde{g}_{\mu\nu}=\mathrm{diag}\{-1, a(t)^2, a(t)^2, a(t)^2\}$, we have for the first Friedman equation:
$H=\frac{\dot{a}}{a}=H_0\sqrt{\Omega_m a^{-3} +\Omega_{rad} a^{-4}+\Omega_\Lambda}$

Here $H=\dot{a}(t)/a$ is the Hubble parameter and $a(t)$ is the scale factor parametrizing the expansion of the Universe. $H_0$ is the current Hubble constant, $\Omega_m$ is the critical matter density (dark matter and baryonic matter), $\Omega_{rad}$ is the critical radiation density, and $\Omega_\Lambda$ is the critical density of the cosmological constant (i.e. dark energy). In our units ($G=1/16\pi$), $\rho_{crit}=6H_0^2$, therefore $\Omega_x=\rho_x/\rho_{crit}=\rho_x/(6H_0^2)$ for $X=\{m,rad,\Lambda\}$.

While the $\Lambda-CDM$ model offers a rather simple explanation of the evolution of the Universe (the minimal $\Lambda-CDM$ has only 6 parameters), it still has its problems. Some of the oldest ones -- the horizon problem, the flatness problem, the missing monopols problem and the large-structures formation problem, require the introduction of a new stage of the development of the Universe -- the inflation. The inflation is an exponential expansion of the Universe lasting between $10^{-36}s$ and $10^{-32}s$ after the Bing Bang, which however increases the volume of the Universe $10^{70}$ times.

The simplest way to produce inflation \cite{9901124} is to introduce a scalar field $\phi$ which is moving in a potential $V_{infl}(\phi)$. Inflation is generated by the exchange of potential energy for kinetic energy. In this case, the evolution of the Universe will be described by two differential equations:
\begin{align}
&H^2=\frac{8\pi}{3m_{Pl}^2}(V_{infl}(\phi)+\frac{1}{2}\dot{\phi}^2)\\
&\ddot{\phi}+3H\dot{\phi}+V_{infl}'(\phi)=0,
\label{st_inf}
\end{align}

\noindent where the first one is the Friedman equation and the second is the inflaton equation. Inflation occurs when $\ddot{a}(t)>0$ which happens in this simple system when $\dot{\phi}^2<V(\phi)$, i.e. when the potential energy dominates over the kinetic one. The pressure and the energy density are: $$p_\phi=\dot{\phi}^2/2-V_{infl}(\phi), \rho_\phi=\dot{\phi}^2/2+V_{infl}(\phi).$$

One can consider different forms for the effective potential, but those simplistic inflationary theories have the problem of not being able to reproduce the graceful transition from inflation to the other observed epochs. 

\section{The multimeasure model} 
There are different ways to obtain a model with richer structure. Here we follow the model developed by Guendelman, Nissimov and Pacheva \cite{ref01, ref01_3, 1408.5344, 1505.07680, 1507.08878, 1603.06231, 1609.06915} (also for some more recent applications of the model \cite{benisty}). The idea is to couple two scalar fields (the inflaton $\phi$ and the darkon $u$) to both standard Riemannian metric and to another non-Riemannian volume form, so that the model can describe simultaneously early inflation, the smooth exit to modern times, and the existence of dark matter and dark energy. 

The action of the model: $S=S_{darkon}\!+\!S_{inflaton}$ is (for more details \cite{1609.06915, 1906.08516}):

$$S_{darkon}=\int d^4x(\sqrt{-g}+\Phi(C))L(u,X_u)$$
$$S_{inflaton}=\int d^4x \Phi_1(A)(R+L^{(1)})+\int d^4x\Phi_2(B)\left(L^{(2)}+\frac{\Phi(H)}{\sqrt{-g}}\right)$$
where $\Phi_i(Z)=\frac{1}{3\!}\epsilon^{\mu\nu\kappa\lambda}\partial_\mu Z_{\nu\kappa\lambda}$ for $Z=A,B,C,H$, are the non-Riemannian measures, constructed with the help of 4 auxiliary completely antisymmetric rank-3 tensors   and we have the following Lagrangians for the two scalar fields $u$ and $\phi$:
\begin{align*}
&L(u)=-X_u -W(u)\\
& L^{(1)}=-X_\phi- V(\phi),  \; V(\phi)=f_1 e^{-\alpha \phi}\\
& L^{(2)}=-{b_0}e^{-\alpha\phi} X_\phi + U(\phi),\; U(\phi)=f_2 e^{-2\alpha \phi}
\end{align*}
\noindent where $X_c=\frac{1}{2}g^{\mu\nu}\partial_\mu c\partial_\nu c$ are the standard kinetic terms for $c=u,\phi$.

Trough the use of variational principle, for this model, it has been found that there exists a transformation \ba 
\tilde{g}_{\mu\nu}=\frac{\Phi(A)}{\sqrt{-g}} g_{\mu\nu}&\\
\frac{\partial\tilde{u}}{\partial u}=(W(u)-2M_0)^{-\frac{1}{2}},&
\ea

for which for the Weyl-rescaled metric $\tilde{g}$, the action becomes
\begin{equation}
S^{(eff)}=\int{d^4x \;\sqrt{-\tilde{g}}(\tilde{R}+L^{(eff)})}\label{sef}.
\end{equation}

For the rescaled metric $\tilde{g}$ and the derived effective Lagrangian, $L_{eff}$, the Einstein Field equations are satisfied.

{\bf The action in the FLRW metric} becomes ($v=\dot{u}$):
\begin{align*}
S^{(eff)} = \int dt \;a(t)^3\Big(-6\;\frac{\dot{a}(t)^2}{a(t)^2} +\frac{\dot{\phi}^2}{2} -\frac{v^2}{2}&\left( V+M_1-\chi_2\eb\dot{\phi}^2/2\right) \\
&\Big. \Big.\Big.\Big.\Big.+\frac{v^4}{4}\left(\chi_2(U+M_2)-2M_0\right)\Big).
 \end{align*}
 from which one can obtain the equations of motion in the standard way.

Explicitly, the equations of motion are:        
\begin{align}
&v^3+3\mba v+2\mbb= 0 \label{sys1}\\
&\dot{a}(t)=\sqrt{\frac{\rho}{6}}a(t), \;\; \label{sys2}\\
&\frac{d}{dt}\left( a(t)^3\dphi(1+\frac{\chi_2}{2}\eb v^2) \right)+a(t)^3 (\alpha\frac{\dphi^2}{2}\chi_2\eb +V_\phi-\chi_2 U_\phi\frac{v^2}{2})\frac{v^2}{2}=0\label{sys3}
\end{align}

Here $\mba_{}=\frac{-1}{3}\frac{V(\phi)+M_1-\frac{1}{2}\chi_2 b e^{-\alpha\phi}\dot{\phi}^2}{\chi_2(U(\phi)+M_2)-2M_0}, \mbb_{}= \frac{-p_u}{2a(t)^3(\chi_2(U(\phi)+M_2)-2M_0)}$ and 
 $$\rho=\frac{1}{2}\dphi^2 (1+\frac{3}{4}\chi_2 b e^{-\alpha\phi} v^2)+\frac{v^2}{4} (V+M_1)+\frac{3 p_u v}{4a(t)^3}$$ is the energy density.

\section{The numerical solutions} 
One can see that the parameters of this system are 12: 4 free parameters $\{\alpha, b_0, f_1, f_2\}$, 5 integration constants $\{M_0, M_1, M_2, \chi_2, p_u\}$ and 3 initial conditions  $\{a(0), \phi(0), \dot \phi (0)\}$. 

We use the following initial conditions:
\ba
a(0)=10^{-10}, \phi(0)=\phi_0, \dot{\phi}(0)=0. 
\label{ini}
\ea

To narrow down the parameter-space, we add also $\{a(1)=1, \ddot{a}(0.71)=0\}$. The consequences of these choices are as follow: 

1) The initial condition $a(0)=0$ introduces a singularity at the beginning of the evolution. 

2) The normalization $a(1)=1$ fixes the age of the Universe. 

3) The condition $\ddot{a}(0.71)=0$ sets the end of the matter-domination epoch.

Defined like this, we have an initial value problem (Eqs.\ref{ini}), which we solve using the shooting method, starting the integration from $t=0$. 

It is possible to also start the integration backwards, from $t=1$, using as initial conditions: $a(1)=1, \phi(1)=\phi_{end}, \dot{\phi}(1)=0$ and aim for $a(0)=0$. Here $\dot{\phi}(1)=0$ guarantees that the evolution of the inflaton field has stopped and the universe is expanding in an accelerated fashion. While both approaches work, integrating forward has the benefit of dealing with the singularity at $a(0)=0$ at the beginning of the integration, rather than at its end. Moving our initial point of integration away from $a(0)=0$ decreases the significance of the term $p_u/a(t)^3$. This effectively means putting $p_u=0$, which we do not want, because $p_u$ is the  conserved  Noether  charge  of the ``dust'' dark matter current (see \cite{1609.06915}). 

The initial velocity of the scalar field $\dot{\phi}(0)$ is not a free parameter of the system, because its value is quickly fixed by the inflaton equation, i.e the results do not depend on $\dot{\phi}(0)$ in a very large interval. 

An important feature of the model, is that the type of evolution one would obtain, depends critically on the starting position on the effective potential. We consider as physically ``realistic'' only the evolution with four epochs ---  short first deceleration epoch (FD), early inflation (EI), second deceleration (SD) which we interpret as radiation and matter dominated epochs together and finally --- slowly accelerating expansion (AE). In terms of the equation of state(EOS) parameter $w(t)=p/\rho$, those are solutions for which: 1) $w_{FD}\to 1/3$, corresponding to the EOS of ultra-relativistic matter, 2) $w_{EI} \to -1$ -- EOS of dark energy, 3) $w_{SD}>- 1/3$ -- EOS of matter-radiation domination, 4) $w_{AE}<-1/3$ -- accelerating expansion period. One obtains this type of solution only for specific choice of the parameters and when starting on the slope of the effective potential. Starting anywhere else results in a non-physical solution (with less epochs). Here we will work only with the ``realistic'' solutions.

Numerically, the times of the different epochs are defined by the three points in which second derivative of the scale factor becomes zero, i.e. $\ddot{a}(t_i)=0$ for $t_i=t_{EI}, t_{SD}, t_{AE}$. In the units we use, $t_{\mathrm{SD}} \sim 10^{-50}$ and $t_{\mathrm{AE}} \sim 0.71$. We  have already reported \cite{1806.08199} a study on how the choice of the parameters affects $t_{\mathrm{SD}}$. Here we will discuss some additional features of the model. 

In \cite{1806.08199} we used the parameter $b_0$ to set $t_{\mathrm{AE}}\sim0.71$ and parameter $f_1$ to ensure $a(1)=1$. Changing $f_1$ however changes the effective potential defined by:
\be
U_{eff}(\phi)=\frac{1}{4}\frac{(f_1 e^{-\alpha\phi}+M_1)^2}{\chi_2 (f_2 e^{-2\alpha\phi}+M_2)-2M_0}. 
\ee
thus making it harder to study how the solutions depend on the starting position on the slope. 

In the current article, we will go a different route and we will use $b_0$ to set $t_{\mathrm{AE}}\sim0.71$ and $p_u$ to ensure $a(1)=1$. This will simplify our problem significantly, since now we will have only 3 parameters to consider $\{b_0, p_u, \phi_0\}$. It will also enable us to study how the solutions depend on $\phi_0$, as the effective potential does not depend on $b_0$ and $p_u$. 

Numerically, we work with the following solution: 

$\{\chi_2=1,\;M_0=-0.034,\;M_1=0.8,\;M_2=0.01, \alpha=2.4, f_1=5, f_2=10^{-5}\}$. 

\begin{figure}[!ht]\centering{
       \subfloat{{\includegraphics[width=4cm]{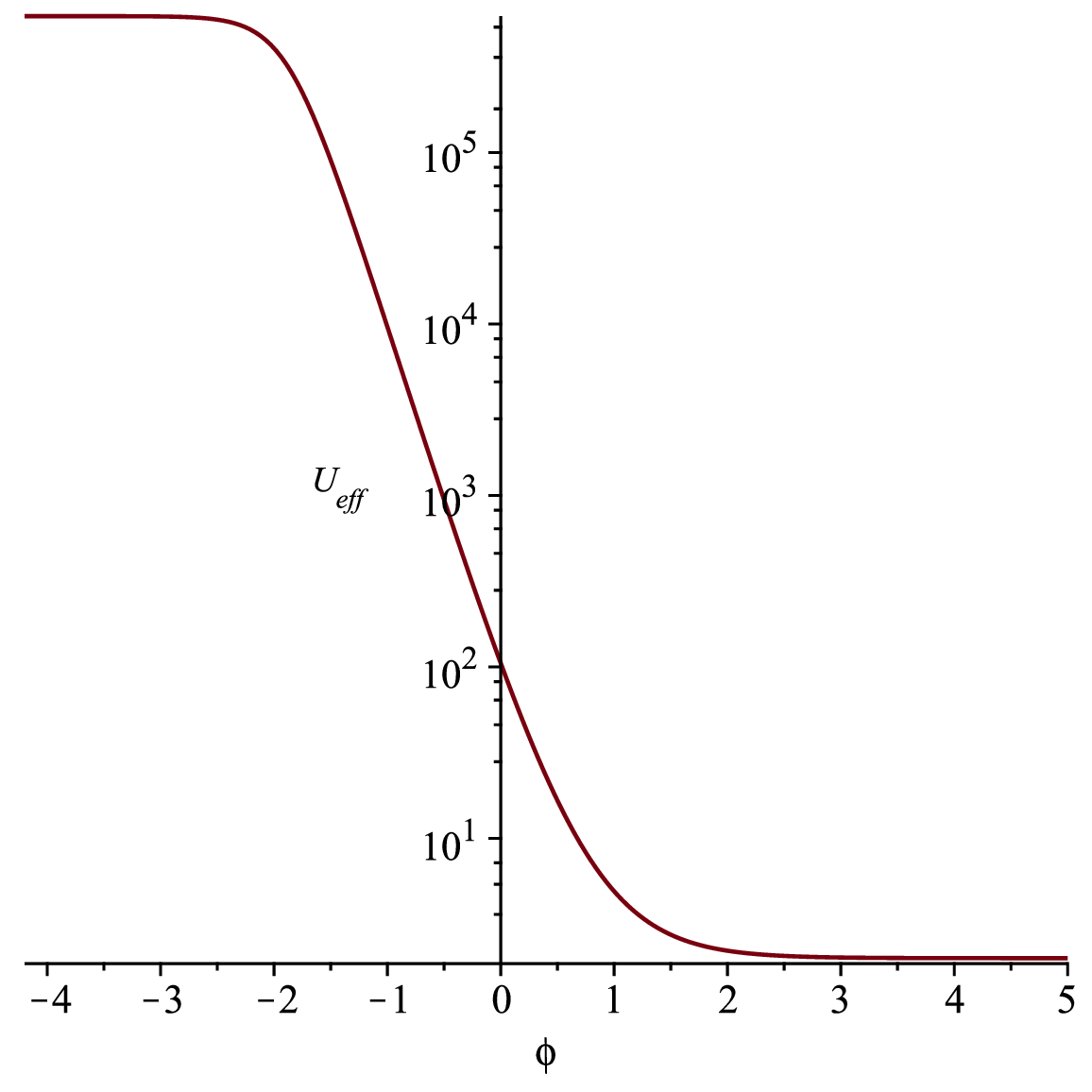} }}
       \subfloat{{\includegraphics[width=4.5cm]{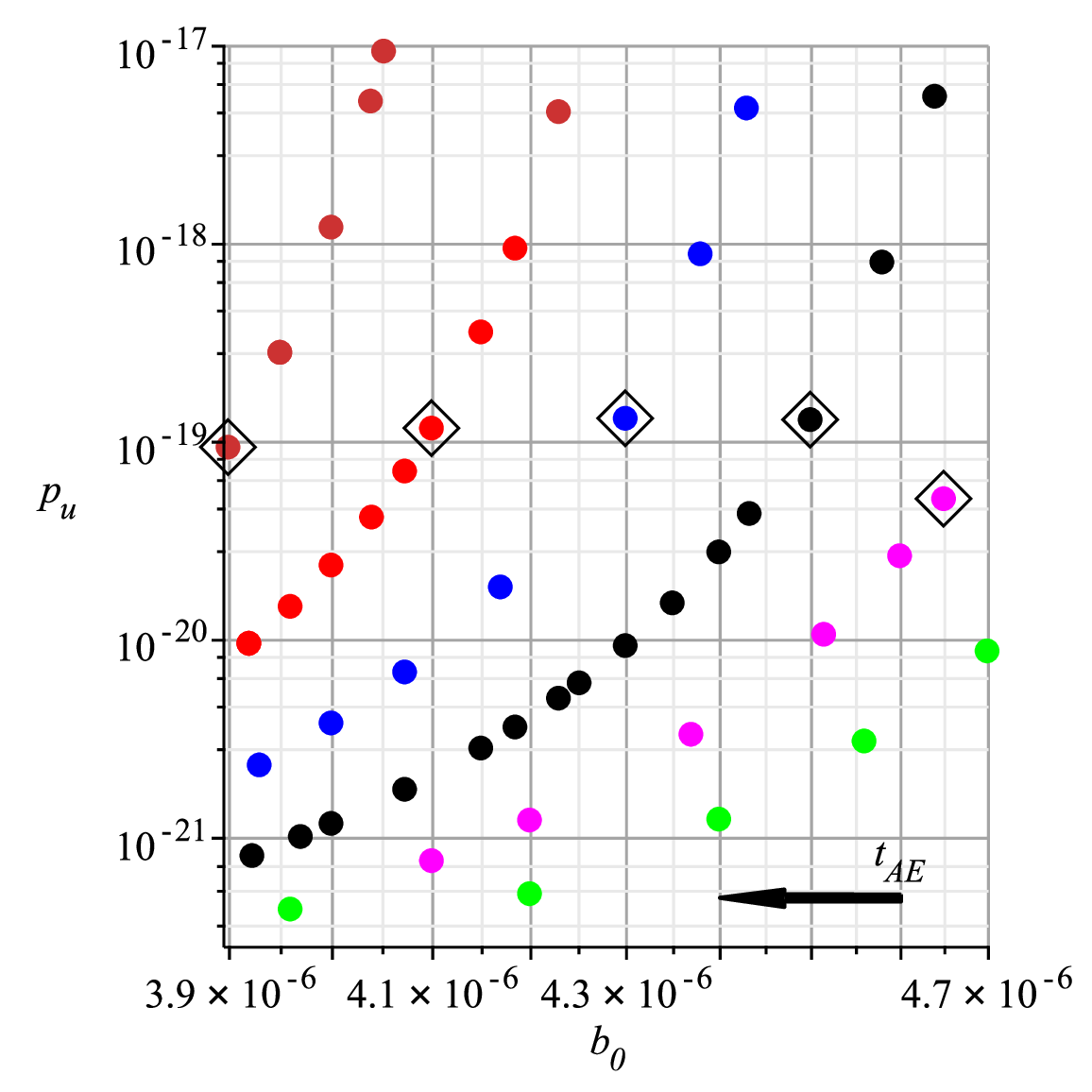} }}\\
       a\hspace{8cm} b\\
        }
   \caption{
 On the panels one can see: 
 (a) the effective potential 
 (d) the plane $p_u(b_0)$, the different branches correspond to different $\phi_0$, the points are solutions, the diamonds are solutions with  $t_{AE}=0.709-0.711$.
}  
   \label{Fig1}
     \end{figure}

For these values of the parameters, the effective potential is step-like, as seen on Fig.\ref{Fig1} a) . The effective potential reaches its asymptotic values for the plateaus for $\phi_-<-4.5, \phi_+>1.7$ (i.e. where  $U_{eff}'\to0$). The slope can be defined as the region $\phi\in(-2.5,-1.2)$, with an inflexion point at $\phi_0=-1.87$.

For this effective potential, one can find different sets of solutions. A family of solutions is shown on Fig.\ref{Fig1} b). One can see the different branches of the solutions corresponding to different $\phi_0$ (in this case $\phi_0\in[-1.5,-1.45]$). As we have mentioned, we require from our solutions to fulfill  $a''(0.71)=0$, i.e. $t_{AE}=0.71$. The points on Fig. 1b) do that with precision of $10^{-2}$, the diamonds show the points which satisfy it with precision of $10^{-3}$ and the different branches correspond to different $\phi_0$. From this plot, one can gain insight on the dependence of $b_0(p_u)$ for the solutions (here different points on the branches correspond to $t_{AE}=0.705..0.715$) and how solutions corresponding to different $\phi_0$ depend on $t_{AE}$ which increases with the decrease of $b_0$ along each branch. 

\begin{figure}[!ht]\centering{
       \subfloat{{\includegraphics[width=4.5cm]{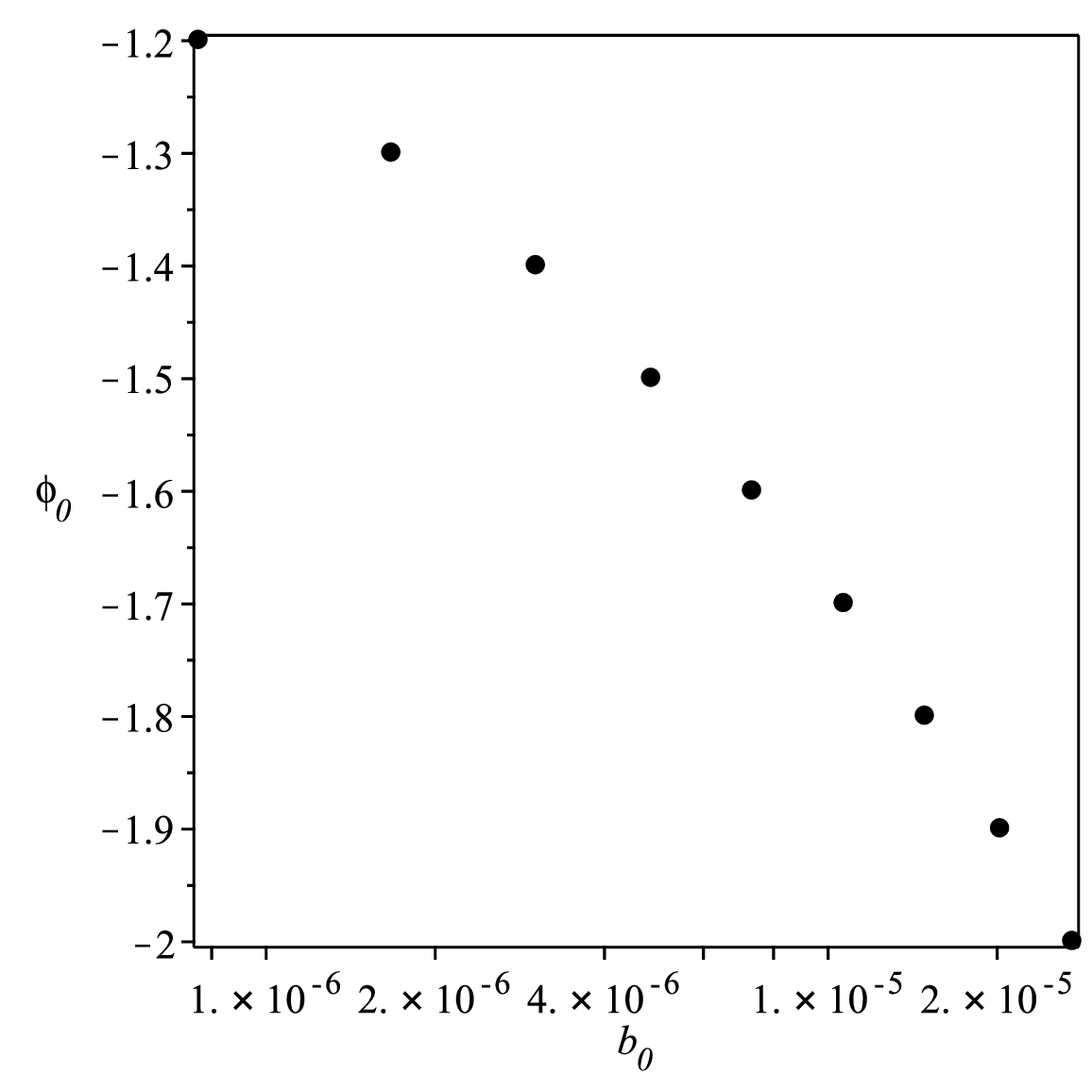} }}
        \subfloat{{\includegraphics[width=4.5cm]{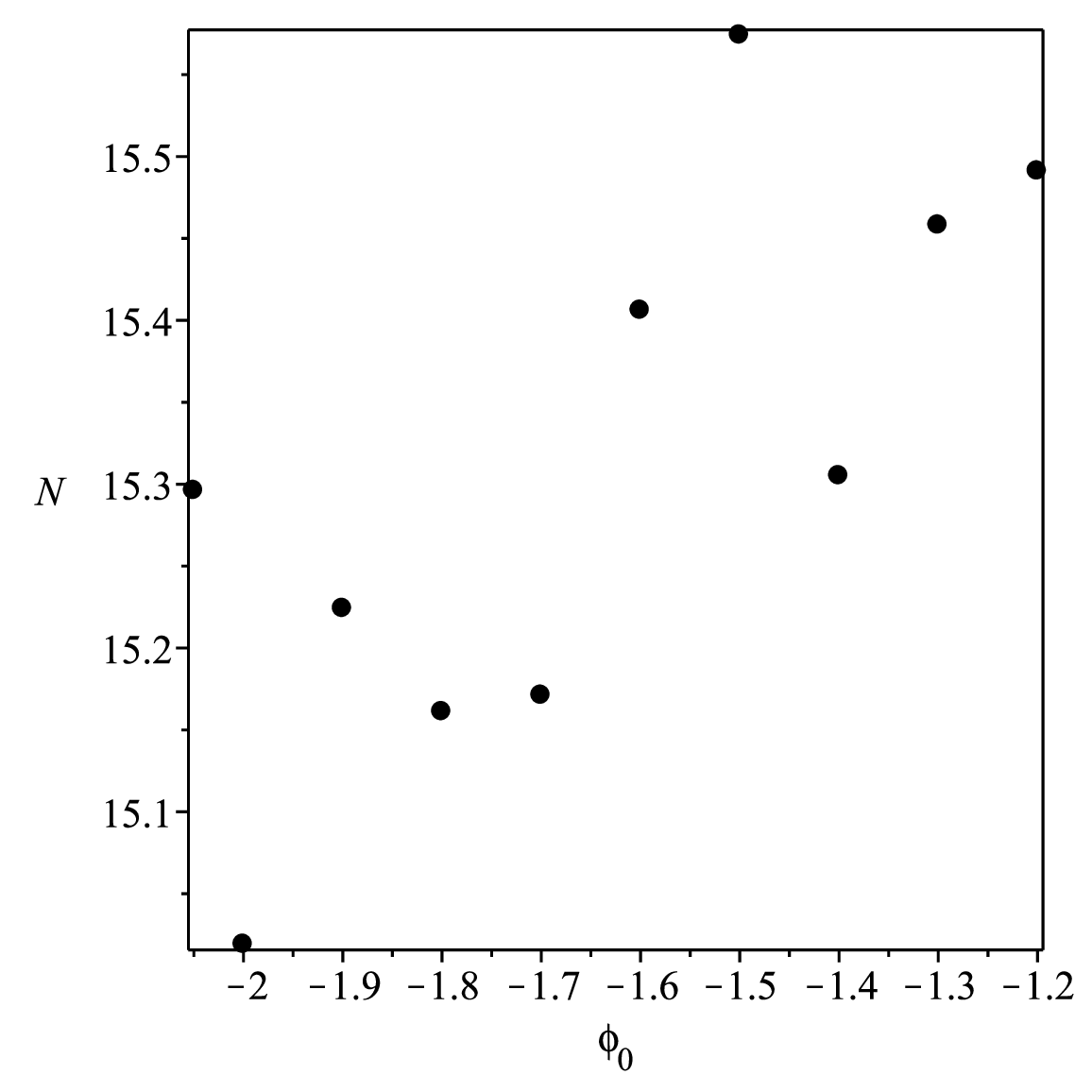} }}\\
       a\hspace{8cm} b\\
        }
   \caption{
 On the panels is 
 (a) the dependence $\phi_0(b_0)$
 (b) the dependence $N(\phi_0)$ 
}  
   \label{Fig2}
     \end{figure}

On Fig. \ref{Fig2} a) we show the dependence $\phi_0(b_0)$ for solutions with $t_{AE}=0.71$ with precision of $10^{-3}$. We do not show the dependence $\phi_0(p_u)$ as it appears chaotic. 

A very interesting feature of these solutions, plotted on Fig.\ref{Fig2} a) and b), is that they keep $t_{EI}$ and $t_{SD}$ approximately constant. I.e. while moving up and down the slope, we do not change the physical properties of our solutions. This can be seen on Fig. \ref{Fig2} b), where we have plotted the e-folds parameter defined as $N=\ln{(a_{SD}/a_{EI})}$ for the solutions corresponding to different $\phi_0$. One can see that it remains more or less the same under the precision we are working with.  In our results, only $\phi(t)$ is sensitive to the changes in $\phi_0$. 

\begin{figure}[!ht]\centering{
       \subfloat{{\includegraphics[width=4.5cm]{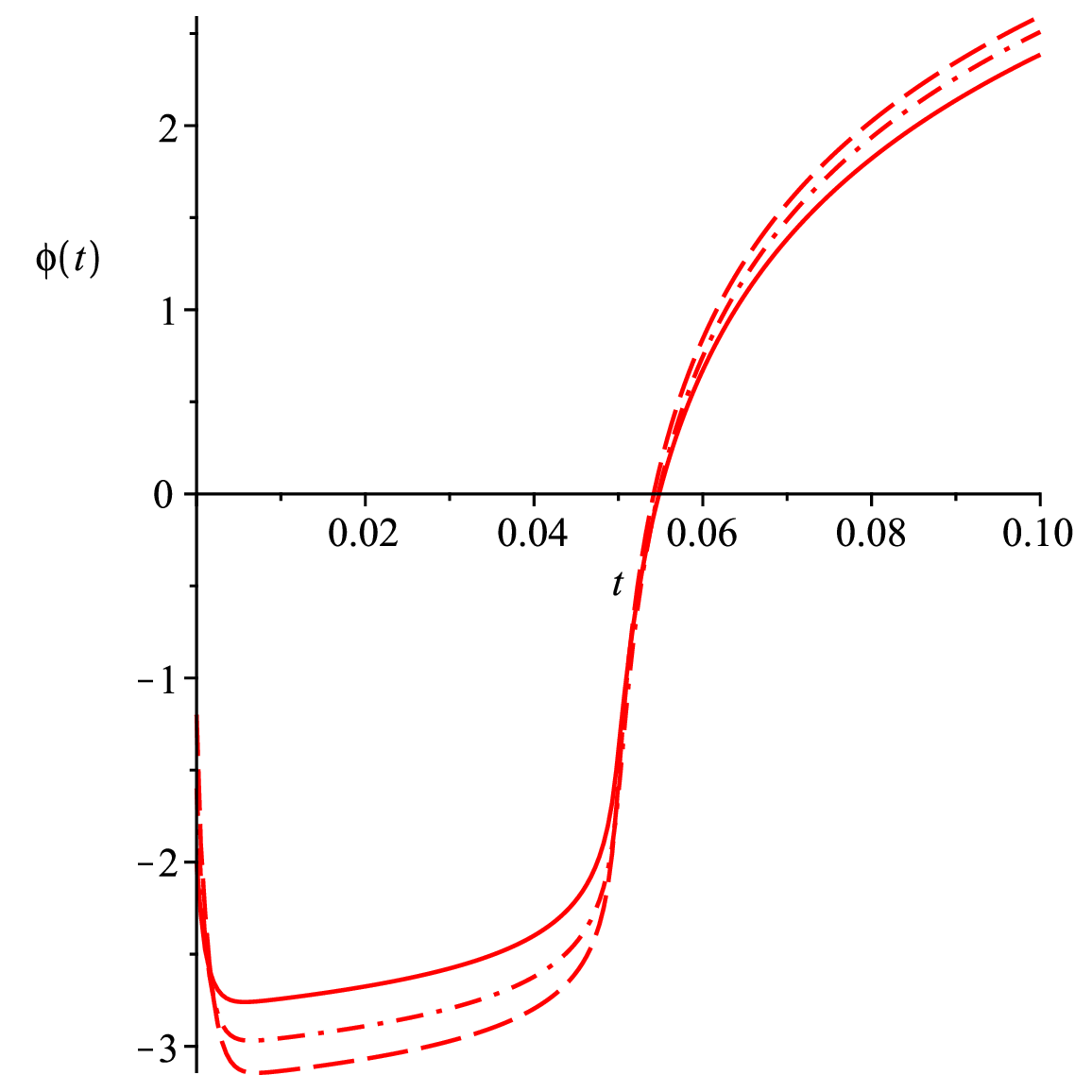} }}
       \subfloat{{\includegraphics[width=4.5cm]{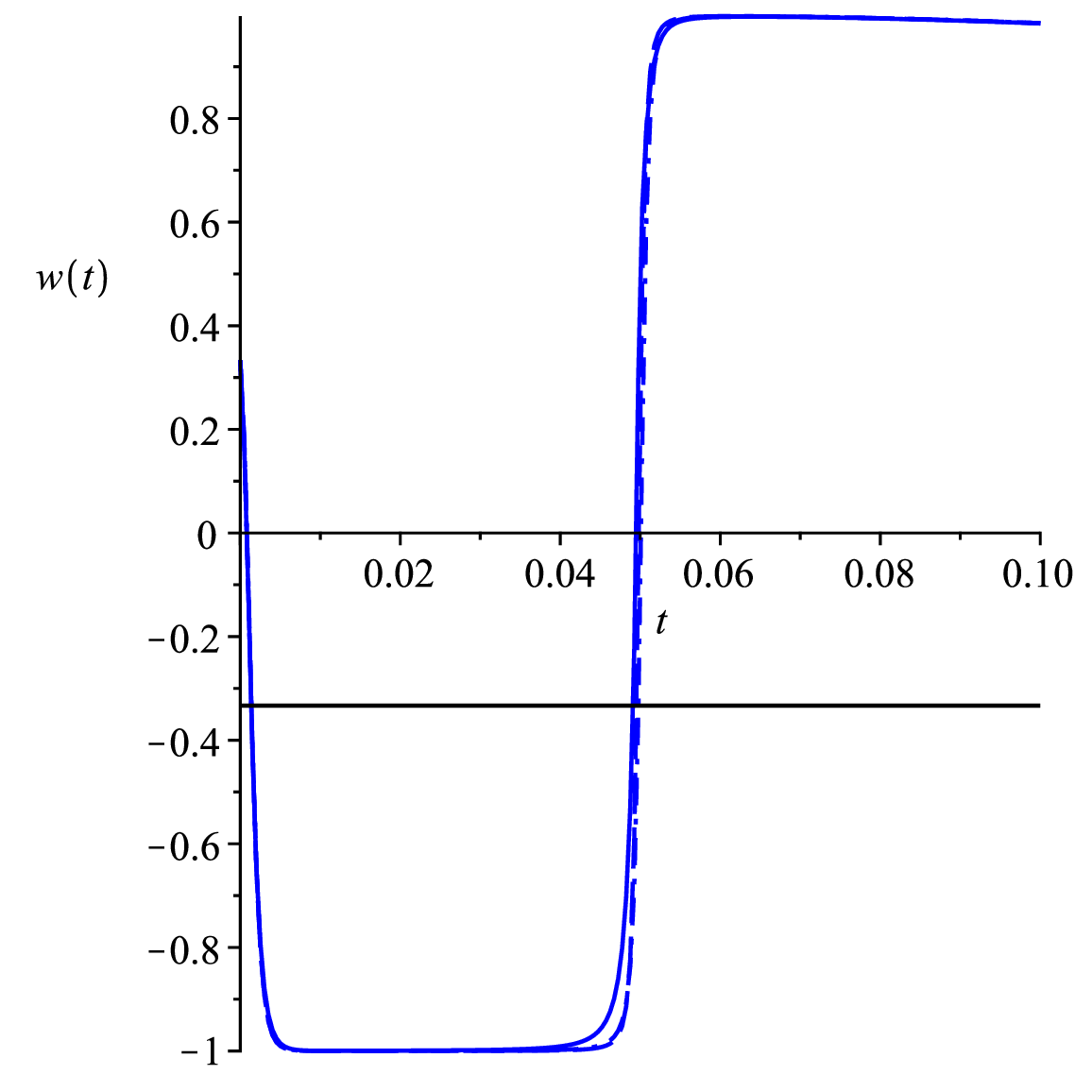} }}\\
       a\hspace{8cm} b\\
        }
   \caption{
Zoomed at $t\in(0,0.1)$ are the evolutions of:
 (a) the inflaton field $\phi(t)$, 
 (b) the equation of state $w(t)=p(t)/\rho(t)$. Here $\phi_0=-1.2, -1.6, -2$ are with dashed, dashed-dotted and solid lines respectively. 
}  
   \label{Fig3}
     \end{figure}

On Fig. \ref{Fig3} a) and b)  we have plotted how $w(t)$ and $\phi(t)$ vary for $\phi_0=\{-1.2, -1.6, -2\}$ and we have zoomed on the interval $t=0..0.1$ where inflation should take place. One can see that there is very little difference in the times of inflation (corresponding to $w(t)=-1$), but there is more visible difference in the evolution of $\phi(t)$. This, however leads to a very small deviation in $\phi(t=1)$ -- less than $5\%$ from the lowest to the highest point on the slope. This result is highly unexpected since one could expect that the power of the inflation and end values of $a(t), \phi(t)$ will depend more strongly on $\phi_0$, which we do not observe here. Note, we could not integrate further up than $\phi_0=-2.05$ which is far from the upper end of the slope $\phi=-2.5$. This is because after this point, $\ddot{\phi}(t)$ becomes infinite and the numerical system hits a singularity. 

Finally, an important note to make is that on Fig. \ref{Fig3}a) one can see the reported before (\cite{1806.08199,1808.08890})   climbing up the slope. Somewhat unexpectedly, it is strongest for points with lowest $\phi_0$. For them, the highest value of $\phi(t)$ is reaching $\phi=-3.10$ which corresponds to the upper plateau.  This is a further confirmation of our observation in \cite{1806.08199} that the movement of the inflaton doesn't correspond to the classical exchange of potential energy for kinetic one, but instead it is closer to the the stability of the L4 and L5 Lagrange points.

This follows from the fact that the effective potential does not bring the kinetic energy in standard form (Eq. \ref{st_inf}). 
In the slow roll approximation (neglecting the terms $\sim\;\dot{\phi}^2,\;\; \dot{\phi}^3,\;\;  \dot{\phi}^4$) the inflaton equation has the form: 
\be 
(A+1)\ddot{\phi}+3H(A+1)\dot{\phi}+U_{eff}'=0,
\label{eq_infl}
\ee 
where $A=\frac{1}{2}b_0e^{-\alpha \phi}\frac{V+M_1}{U+M_2}$.

For this reason, we find it interesting to compare the kinetic slow-roll parameters (also called dynamical) defined as:
\be
\epsilon_H=-\frac{\dot{H}}{H^2},\;\; \eta_H=-\frac{\ddot{\phi}}{H\dot{\phi}}
\label{eps_eta}
\ee
\noindent 
with the potential slow-roll parameters which can be derived to be \cite{9408015,1408.5344} (Note that our $A$ is different from the one used in \cite{1408.5344}): 
\be
\epsilon_V=\frac{1}{1+A}\left(\frac{U_{eff}'}{U_{eff}}\right)^2,\;\; \eta_V=\frac{2}{1+A} \frac{U_{eff}''}{U_{eff}}
\label{eps_eta_v}
\ee

 \begin{figure}[!h]\centering{
      \subfloat{{\includegraphics[width=4cm]{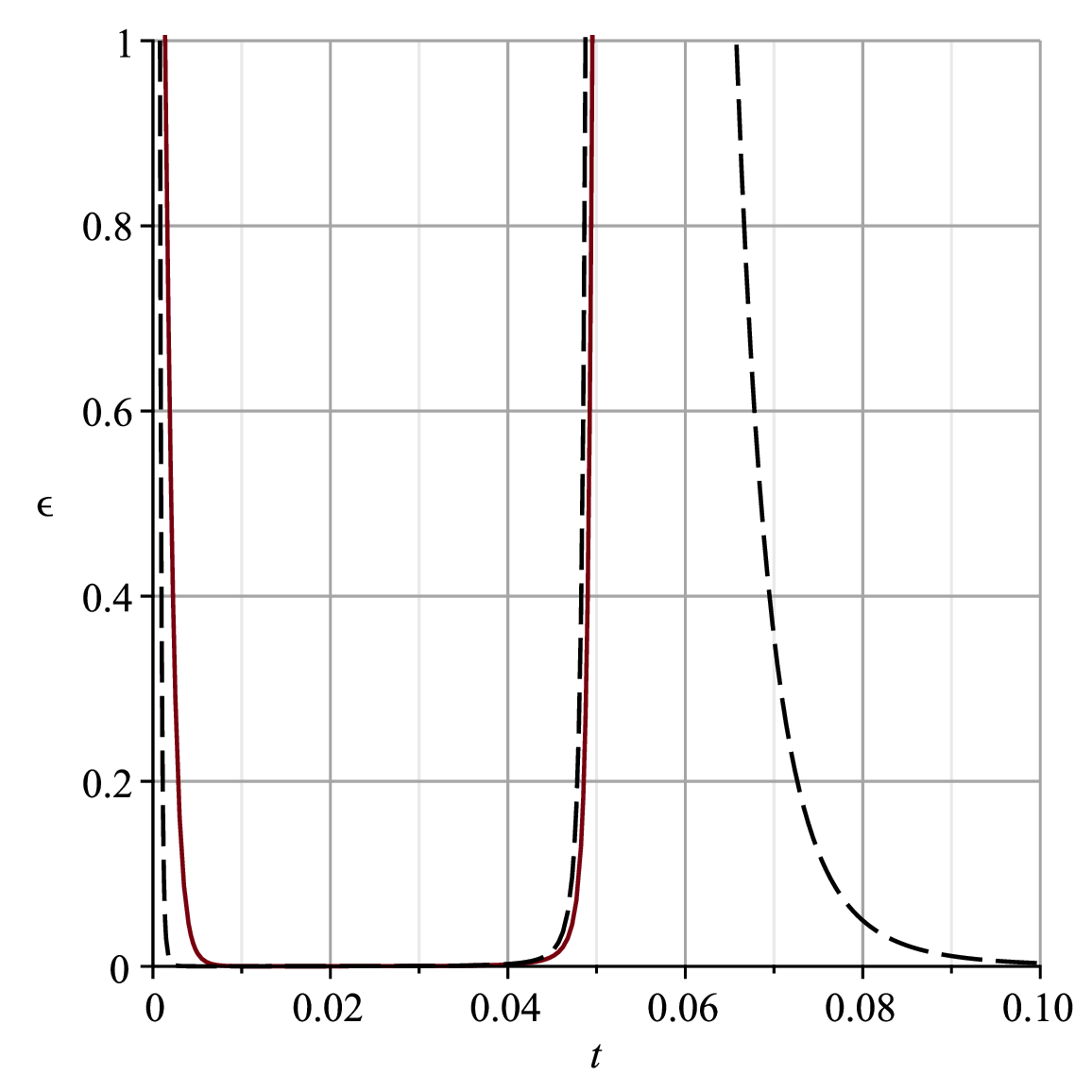} }}
       \subfloat{{\includegraphics[width=4cm]{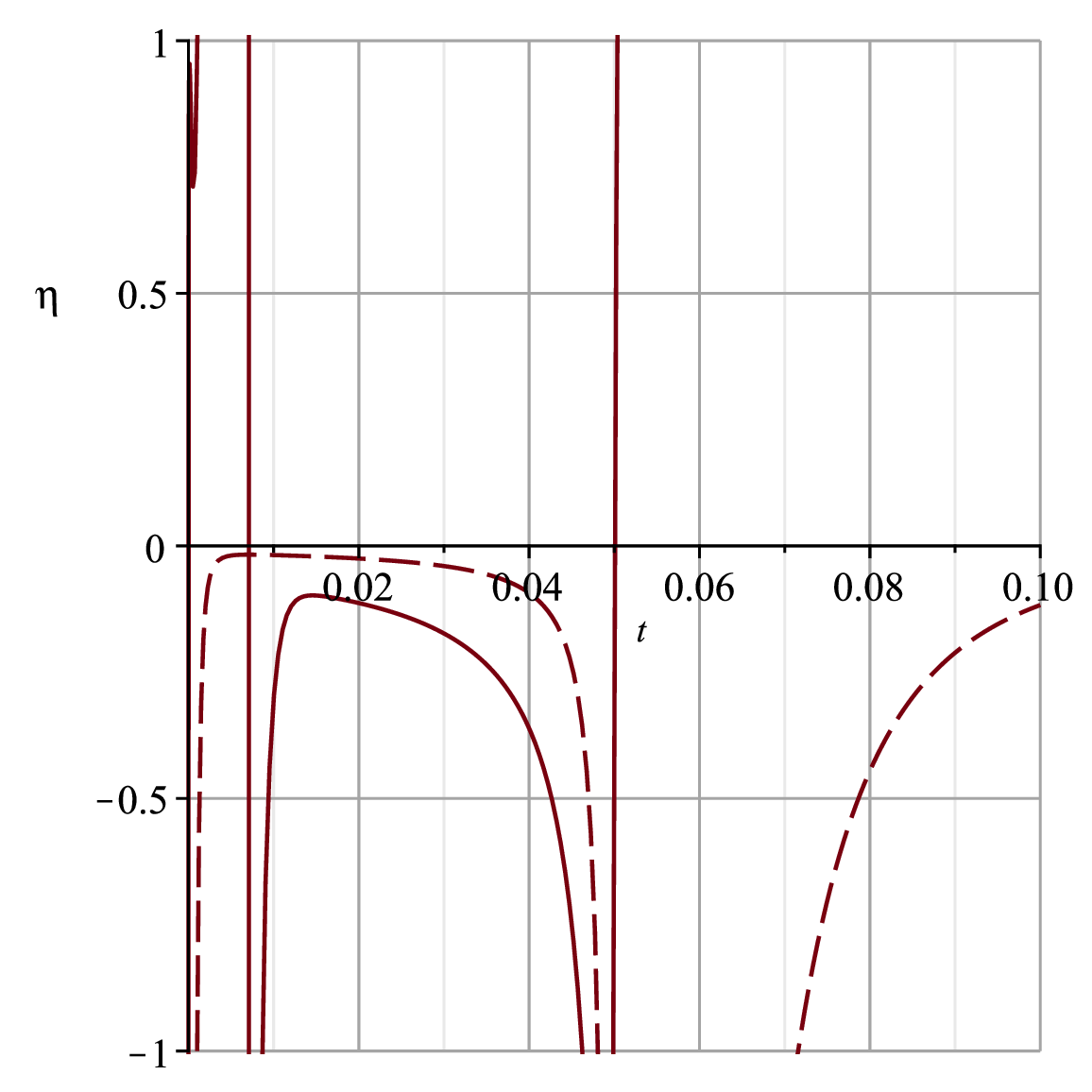} }}\\
    a\hspace{4cm} b}
    \caption{A zoom in on the slow-roll parameters ((a) $\epsilon$ and (b) $\eta$) in the interval where infation occurs $(t=0.00138..0.04953)$. With a solid black line are plotted the dynamical parameters, with the black dashed one -- the potential ones. The values of $\{b_0,p_u,\phi_0\}$ are $\{0.76\times 10^{-6},0.53\times 10^{-19},-1.2\}$}  
   \label{Fig4}
     \end{figure}

On Fig. \ref{Fig4} we present an example of the evolution of the ``slow-roll'' parameters for both the kinetic and the potential definitions \cite{9408015,1408.5344}. 
One can see from the plots that the two definitions in this interval are very similar  -- both give mostly very small values of the slow-roll parameters. One also notices that the intervals on $t$ for which the slow-roll parameters of both kinds are very small (say, $|<0.1|$) are shorter than the numerically obtained one, given by $t_{EI}..t_{SD}$ \footnote{To be precise, the intervals are:
for $\epsilon_V: 0.0012..0..0473$, 
for $\epsilon_H: 0.0034..0.0481$
for $\eta_V: =0.0021..0.041$
for $\eta_H: 0.013..0.033$
}. 
In general, it seems that the potential slow-roll parameters give intervals closer to the numerical ones. However, the conclusion is that if the slow-roll parameters are used to estimate inflation theoretically, those small deviations in the intervals may lead to misestimations of $N$. 

Finally, a note on the e-folds parameter, which measures the power of the inflation. The theoretical estimation for the number of e-folds needed to solve the horizon problem is model-dependent but is $N>70$. In our example, we get $N\approx 15$. It is important to note, that there is a numerical maximum of the number of e-folds of about $N\approx22$, due to the fact we are starting our integration at $a(0)\sim 10^{-10}$. In order to get a higher $N$, one needs to start with smaller $a(0)$, but to do so, we need to improve significantly the precision of the integration.  Parameters-wise, the best way to get powerful early inflation is trough increasing $\alpha$ or decreasing $f_2$ \cite{1806.08199}.

\section{Conclusions}
In this work, we have explored numerically the model of Guendelman-Nissimov-Pacheva in a specific part of its parametric space related to the different initial conditions on the slope of the effective potential. 
Even though it is impossible to study the entire parameter space, we have shown some important properties of the model. 

Most importantly, we have shown that there exist families of solutions on the slope which preserve the initial and the ending times of the inflation and also, that they give similar number of e-folds. 
Furthermore, one can see that the main difference between starting at the top of the slope and at its bottom is the behavior of the inflaton scalar field, which climbs up all the way to the top of the slope before entering in  inflation regime. This mechanism requires further study. 

Finally, we have considered the dynamical and the potential slow-roll parameters for the model and we have shown that the potential slow-roll parameters seem to describe the inflationary period better. They, however, do not match entirely with the numerically obtained duration of the inflation. 

\section*{Acknowledgments}

It is a pleasure to thank E. Nissimov, S. Pacheva and M. Stoilov for the discussions. 

The work is supported by Bulgarian NSF grant DN-18/1/10.12.2017, KP-06-N 38/11 and by Bulgarian NSF grant 8-17. D.S. is also partially supported by COST Actions CA18108.

\end{document}